\documentclass[nofootinbib,12pt]{article}
\usepackage{graphicx}

\begin{document}

\parskip=0.3cm


\vskip 0.5cm \centerline{\bf Indirect evidence of the Odderon
 from the LHC data on elastic proton-proton scattering
} \vskip 0.3cm

\centerline{  A.I.~Lengyel $^{a}$, Z.Z. Tarics $^{b}$}

\vskip 1cm

 \centerline{ \sl Institute of Electron
Physics, Nat. Ac. Sc. of Ukraine, Uzhgorod}
\vskip
0.1cm

\vskip 1cm

$
\begin{array}{ll}
^{a}\mbox{{\it e-mail address:}} & \mbox{alexander-lengyel@rambler.ru}\\

^{b}\mbox{{\it e-mail address:}} & \mbox {tarics1@iep.uzhgorod.ua}\\
\end{array}
$

\begin{abstract}
A simple multipole Pomeron and Odderon model for elastic hadron scattering,
reproducing the structure of the first and second diffraction cones is used to analyze $pp$ and $\overline{p}p$ scattering.
The main emphasis is on the delicate and
non-trivial dynamics in the dip-bump region, at $\left|t\right| \approx  1$ GeV$^2$ and at the second cone.
The simplicity of the model and the expected smallness of the
absorption corrections enables one the control of various
contributions to the scattering amplitude, in particular the
interplay between the C-even and C-odd components of the
amplitude, as well as their relative contribution, changing with
$s$ and $t$. The role of the non-linearity of the Regge
trajectories is verified.
A detailed analysis of the LHC energy region, where most of the exiting
models may be either confirmed or ruled out, is presented.

\end{abstract}

\vskip 0.1cm

\section{Introduction} \label{s0}

 The long-standing debate about the existence of the Odderon ($C$-odd partner of the Pomeron) can be resolved definitely only by a high-energy experiment involving particle and anti-particle scattering, e.g.
$pp$ and $\bar pp$ scattering, in the same kinematical region. There was a single experiment of that kind, at the ISR ~\cite{ISR}, where the two cross sections were found to differ. The unique observation, however relies on a few data points only, and ISR was shut down shortly after that experiment, leaving some doubts on the validity of the
effect. Moreover, the ISR energies were not high enough to exclude the alternative explanation of the difference, namely due to $\omega$ exchange still noticeable at the ISR in the region of the dip. This is not the case at the LHC, where the contribution from secondary trajectories can be practically excluded within the diffraction cone region. Waiting for a possible future upgrade of the LHC energy down to that of the Tevatron, which will enable
a direct confrontation of $pp$ and $\bar pp$ data, here we analyze the available LHC data on $pp$ scattering in a model with and without the Odderon contribution. Anticipating the final result, let us mention that one should not
dramatize the question of the (non)existence of the Odderon: in our opinion, it exists simply because nothing forbids its existence. The only question is its parametrization and/or relative contribution with respect e.g. to the Pomeron.
Due to the recent experiments on elastic and inelastic proton-proton scattering by the TOTEM Collaboration at the LHC ~\cite{7pp}, data in a wide range, from lowest up to TeV energies,  both for proton-proton and antiproton-proton scattering in a wide span of transferred momenta are now available.
The experiments at TeV energies gives an opportunity to verify different Pomeron and Odderon models because the secondary Reggeon contributions at these energies are small.
However none of the existing models of elastic scattering was able to predict the value of the differential cross section beyond the first cone, as clearly seen in Fig.4 of the TOTEM paper \cite{7pp}.

It should be noted that the predictions of Regge-pole models are rather qualitative, so the new experimental data always stimulate their improvement.
Let us remind that the ISR  measurements stimulated the development of multipole Pomeron models, including the dipole one, that successively described the dip - bump structure and both cones of the differential cross section of hadron-hadron scattering ~\cite{VJS}.

The first attempt to describe high-energy diffraction peculiarities
in the differential cross sections, was  made by Chou and Yang
in  ``geometrical'' \cite{C-Y} model, which
qualitatively reproduces the $t$ dependence of the
differential cross sections in elastic scattering, however it does
not contain any energy dependence, subsequently introduced by
means of Regge-pole models.
An example to examine the role of dipole Pomeron (DP), we performed the control fit for data of ISR in the model of dipole Pomeron (see below). As  result, we curtained of  that the role of Odderon headily grows with the height of energy.

In recent paper~\cite{JLL} we have used a simple dipole Pomeron model that
reproduces successfully  the structure of first and second diffraction cones in $pp$ and $\overline{p}p$ scattering.
The simplicity and transparency of the model enables one to control of various
contributions to the scattering amplitude, in particular the interplay between the C-even and C-odd components of the amplitude, as well as their relative contribution, changing with
$s$ and $t$.
It was shown that, while the contribution from secondary Reggeons
is negligible at the LHC, the inclusion of the Odderon is
mandatory, even for the description of $pp$ scattering alone.
Therefore the precise measurement of $ pp $ differential cross section gives a chance to distinguish various models of Pomeron  ~\cite{Godizov} and especially Odderon \cite{Nicolescu2}, \cite{Nicolescu}.
To do this one needs to compare the predictions of the models. Such a comparison makes sense only if the same data set is used when the parameters of the models are determined.

The possible extensions of DP model include:
\begin{itemize}
   \item The dip-bump structure typical to high-energy diffractive processes;
   \item Non-linear Regge trajectories;
   \item Possible Odderon (odd-$C$ asymptotic Regge exchange);
   \item Compatible with $s-$ and $t-$ channel unitarity;
\end{itemize}

Below we suggest a simple model that can be used as a handle in studying diffraction
 at the LHC. It combines the simplicity of the above models approach,
 and goes beyond their limitations. Being flexible, it can be modified according to
the experimental needs or theoretical prejudice of its user and
can be considered as the ``minimal model'' of high-energy
scattering while its flexibility gives room for various
generalizations/modifications or further developments (e.g.
unitarization, inclusion of spin degrees of freedom  etc.).
To start with, we choose the model, successfully describing $pp$ and $\overline{p}p$ scattering \cite{JLL} within the framework of the simple dipole Pomeron.
Assuming that the role of the Odderon  in the second cone increases with energy, for more adequate definition of data we vary
the form of the Odderon. Being limited in our choice, we will chose an Odderon copying many features of the Pomeron, e.g. its trajectory being non-linear.

In this paper, we consider the spinless case of the invariant
high-energy scattering amplitude, $A\left(s,t\right)$, where $s$
and $t$ are the usual Mandelstam variables.
The basic assumptions
of the model are:

1. The scattering amplitude is a sum of four terms, two asymptotic
(Pomeron (P) and Odderon (O)) and two non-asymptotic ones or secondary Regge pole contributions.
where  $P$ and $f$ have positive
$C$-parity, thus entering in the scattering amplitude with the
same sign in $pp$ and $\overline{p}p$ scattering, while the Odderon and
$\omega$ have negative $C$-parity, thus entering
$pp$ and $\bar pp$ scattering with opposite signs, as shown below:
 \begin{equation}\label{Eq:Amplitude}
 A\left(s,t\right)_{pp}^{\bar pp}=A_P\left(s,t\right)+
 A_f\left(s,t\right)\pm\left[A_{\omega}\left(s,t\right)+A_O\left(s,t\right)\right],
 \end{equation}
where the symbols $P,\ f,\ O,\  \omega$ stand for the relevant
Regge-pole amplitudes and the super(sub)script, evidently,
indicate $\bar pp (pp)$ scattering with the relevant choice of
the signs in the sum (\ref{Eq:Amplitude}).

 2. We treat the Odderon, the $C$-odd counterpart of the Pomeron on equal
 footing, differing by its $C-$ parity and the values of its parameters
 (to be fitted to the data).
 We examined also a fit to $pp$ scattering alone, without any Odderon contribution. The (negative) result is presented in Sec.~\ref{sec:odderon};

 3. The main subject of our study is the Pomeron  and  the Odderon, as a double poles, or DP \cite{VJS, reviews}) lying on a nonlinear trajectory, whose intercept is not equal to one. This choice
is motivated by the unique properties of the DP: it produces
logarithmically rising total cross sections at unit Pomeron
intercept. By letting $\alpha_P\left(0\right)>1,$ we allow for a
faster rise of the total cross section, although the intercept is about half that in the DL model since the double pole (or dipole) itself drives the rise in energy.
 A supercritical Pomeron trajectory,  $\alpha_P(0)>1$ in the DP is required by the
observed rise of the ratio $\sigma_{el}/\sigma_{tot},$ or,
equivalently, departure form geometrical scaling \cite{VJS}.
The dipole Pomeron produces logarithmically rising total cross
sections and nearly constant ratio of $\sigma_{el}/\sigma_{tot}$ at
unit Pomeron intercept, $\alpha_P\left(0\right)=1.$
In addition this  mild logarithmic increase of $\sigma_{tot}$ does not supported by the
 result of the last experiment at LHC for energy 7 TeV $ \sigma _{tot} = (98.3 \pm 2.8 \pm 0.02) mb $ \cite {7tev}.
Along with the rise of the ratio $\sigma_{el}/\sigma_{tot}$ beyond the SPS
energies requires a supercritical DP intercept,
$\alpha_P\left(0\right)=1+\delta,$ where $\delta$ is a small
parameter $\alpha_P(0)\approx 0.05$. Thus DP is about ``twice softer''
then that of Donnachie-Landshoff \cite{DL}, in which
$\alpha_P(0)\approx 0.08.$
Due to its geometric form (see below) the DP reproduces itself against unitarity (eikonal)
corrections. As a consequence, these corrections are small, and
one can use the model at the ``Born level'' without complicated
(and ambiguous) unitarity (rescattering) corrections. DP combines
the properties of Regge poles and of the geometric approach,
initiated by Chou and Yang, see \cite{C-Y}.

4. Regge trajectories are non-linear complex functions.
This nonlinearity is manifest e.g. as the ``break'' i.e. a change the slope $\Delta B \approx 2$ GeV$^2$ around $t\approx-0.1$ GeV$^2$ and at large $|t|$, beyond the second maximum we observe nonzero curvature at least for wide $-t$ region.
In spite of a great varieties of models for high-energy diffraction (for a recent review see \cite{review}), only a few of them attempted to attack the complicated and delicate mechanism of
the diffraction structure. In the 80-ies and early 90-ies, DP was fitted to the ISR, SPS and Tevatron data, see \cite{reviews, 2,3, KLT} and \cite{VJS} for earlier references. Now we find it appropriate to revise the state of the art in this field, to update the earlier fits,  analyze the ongoing measurements at the LHC and/or make further predictions. We revise the existing estimates of the Pomeron and particularly Odderon contributions to the cross sections as a functions of $s$ and $t$ and argue that while the contribution from non-leading
trajectories in the nearly forward region is negligible (smaller than the experimental uncertainties), the Odderon may be important, especially beyond the first cone.

\section{The model} \label{s2}
\vskip 0.2cm We use the normalization:
\begin{equation}\label{norm}
{d\sigma\over{dt}}={\pi\over s^2}|A(s,t)|^2\ \  {\rm and}\ \
\sigma_{tot}={4\pi\over s}\Im m A(s,t)\Bigl.\Bigr|_{t=0}\ .
\end{equation}
Neglecting spin dependence, the invariant proton(antiproton)-proton elastic scattering amplitude is that of Eq. (\ref{Eq:Amplitude}).
The secondary Reggeons are parametrized in a standard way with linear Regge trajectories and
exponential residua, where $R$ denotes $f$ or $\omega$ - the
principal non-leading contributions to $pp$ or $\bar p p$
scattering:
\begin{equation}\label{Reggeons}
A_R\left(s,t\right)=a_R{\rm e}^{-i\pi\alpha_R\left(t\right)/2}{\rm e}
^{b_Rt}\Bigl(s/s_0\Bigr)^{\alpha_R\left(t\right)},
\end{equation}
with handbook slopes  $\alpha'_f\left(t\right)=0.84$ and
$\alpha'_{\omega}\left(t\right)=0.93.$ The values of other parameters of the Reggeons are quoted in Table~\ref{tab:fitparam}.
As argued in the Introduction, the Pomeron is a dipole in the $j-$plane
\begin{equation}\label{Pomeron}
A_P(s,t)={d\over{d\alpha_P}}\Bigl[{\rm
e}^{-i\pi\alpha_P/2}G(\alpha_P)\Bigl(s/s_0\Bigr)^{\alpha_P}\Bigr]=
\end{equation}
$${\rm
e}^{-i\pi\alpha_P(t)/2}\Bigl(s/s_0\Bigr)^{\alpha_P(t)}\Bigl[G'(\alpha_P)+\Bigl(L-i\pi
/2\Bigr)G(\alpha_P)\Bigr].$$
Since the first term in squared brackets determines the shape of
the cone, one fixes
\begin{equation} \label{residue} G'(\alpha_P)=-a_P{\rm
e}^{b_P[\alpha_P-1]},\end{equation} where $G(\alpha_P)$ is recovered
by integration, and, as a consequence, the Pomeron amplitude Eq.
(\ref{Pomeron}) can be rewritten in the following ``geometrical''
form (for the details of the
calculations see \cite{VJS} and references therein)
\begin{equation}\label{GP}
A_P(s,t)=i{a_P\ s\over{b_P\ s_0}}[r_1^2(s){\rm
e}^{r^2_1(s)[\alpha_P-1]}-\varepsilon_P r_2^2(s){\rm
e}^{r^2_2(s)[\alpha_P-1]}],
\end{equation}
where
\begin{equation}\
r_1^2(s)=b_P+L-i\pi/2,\ \ r_2^2(s)=L-i\pi/2,\ \ L\equiv
\ln(s/s_0).
\end{equation}
We use a representative example of the Pomeron trajectory, namely that with a two-pion square-root threshold, Eq.~(\ref{eq:tr2}), required by $t-$channel unitarity and accounting for the small-$t$ ``break'' \cite{Cohen},
\begin{equation}
\alpha_P\equiv \alpha_P(t) =
1+\delta_P+\alpha_{1P}t - \alpha_{2P}\left(\sqrt{4m_{\pi}^2 -t}-2 m_{\pi}\right),
\label{eq:tr2}
\end{equation}
where $m_{\pi}$ - pion mass.
An important property of the DP Eq.~(\ref{GP}) is the presence of
absorptions, quantified by the value of the parameter
$\varepsilon_P$. This property, together with the non-linear nature of the trajectories, justifies the neglect of the rescattering corrections.  More details can be found e.g.
in Ref. \cite{VJS}.)
The unknown Odderon contribution is assumed to be of the same form
as that of the Pomeron, Eqs.~(\ref{Pomeron}),~(\ref{GP}), apart from different values of adjustable parameters (labeled by the subscript ``$O$''). 
\begin{equation}\label{Odd}
A_O(s,t)={a_O\ s\over{b_O\ s_0}}[r_{1O}^2(s){\rm
e}^{r^2_{1O}(s)[\alpha_O-1]}   -\varepsilon_O r_2^2(s){\rm
e}^{r^2_{20}(s)[\alpha_O-1]}],
\end{equation}
where
\begin{equation}
r_{1O}^2(s)=b_O+L-i\pi/2,\ \ r_{2O}^2(s)=L-i\pi/2,\ \ L\equiv
\ln(s/s_0).
\end{equation}
and
\begin{equation}
\alpha_O\equiv \alpha_O(t) =
1+\delta_O+\alpha_{1O}t - \alpha_{2O}\left(\sqrt{4m_{\pi}^2 -t}-2 m_{\pi}\right),
\label{eq:trOd}
\end{equation}
The form and properties of Odderon trajectory  is the same along with the scale value $s_{0}=100 GeV^{2}$.
The adjustable parameters are: $\delta_P,\ \alpha_{iP},\
a_P,\ b_P,\ \varepsilon_P$ for the Pomeron and
$ \alpha_{iO},\ a_O,\delta_O,\ b_O, \varepsilon_O$ for
the Odderon. The results of the fitting procedure is  presented
below.

\section{Fits without the Odderon}\label{sec:odderon}
\begin{figure}[fig:pomeron]
\center{
        \includegraphics[width=0.45\linewidth]{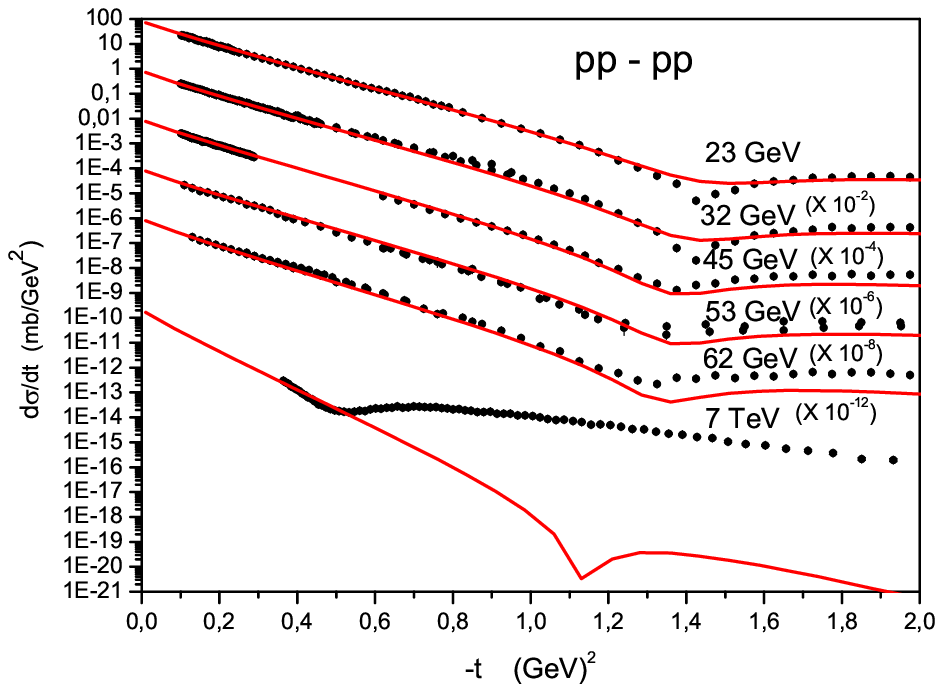}
        \includegraphics[width=0.45\linewidth]{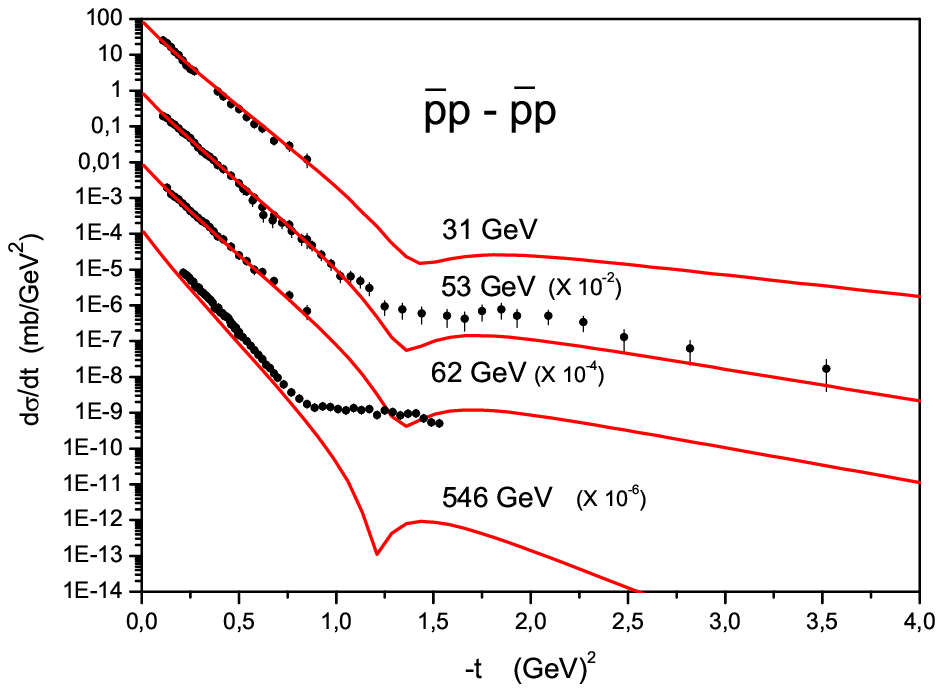}\\
        (a)\hspace{0.45 \linewidth}(b)}
\caption{Differential $pp$ (a) and $\bar{p}p$ (b) cross sections fitted without the Odderon term to the ISR data, calculated from the model of the  previous Section.
\label{fig:pomeron}
}
\end{figure}

To check the role of the Odderon, we first fit only $pp$ scattering without any Odderon (supposed to fill the dip in $\bar pp$). The resulting fit is shown in Fig~\ref{fig:pomeron}, demonstrating that, while the Pomeron appended with sub-leading reggeons  reproduces qualitatively the dip for low  energies, namely 23, 32, 45, 53 and 62 GeV \cite{ISR,Amaldi:1979kd, Albrow:1976sv, Breakstone:1984te, data}. The dipole Pomeron model gives a good description of the first and second cones,
but deteriorates with increasing energy in range of the second cone. It is special notable in the energy inerval $0.5-7$ TeV.
In fig. ~\ref{fig:pomeron}~(b) the  $\bar pp$ differential cross section calculated with the same parameters is shown. Apart for a shoulder instead of the dip in $pp$, the quality of the fit beyond this shoulder is comparable to that in $pp$.

\begin{figure}[fig:pomeron2]
\center{
		\includegraphics[width=0.45\linewidth]{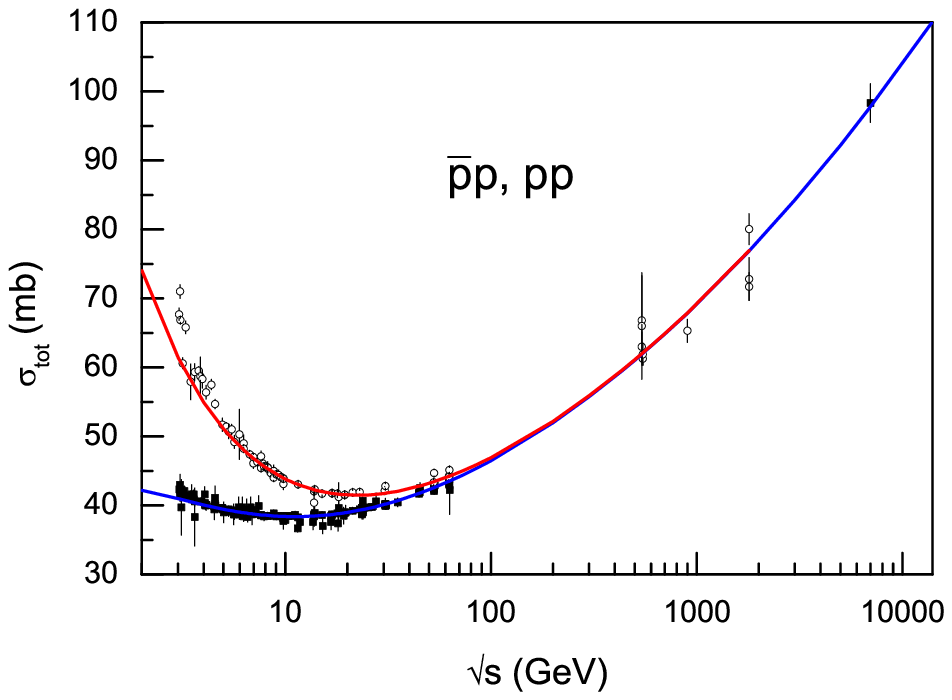}
		\includegraphics[width=0.45\linewidth]{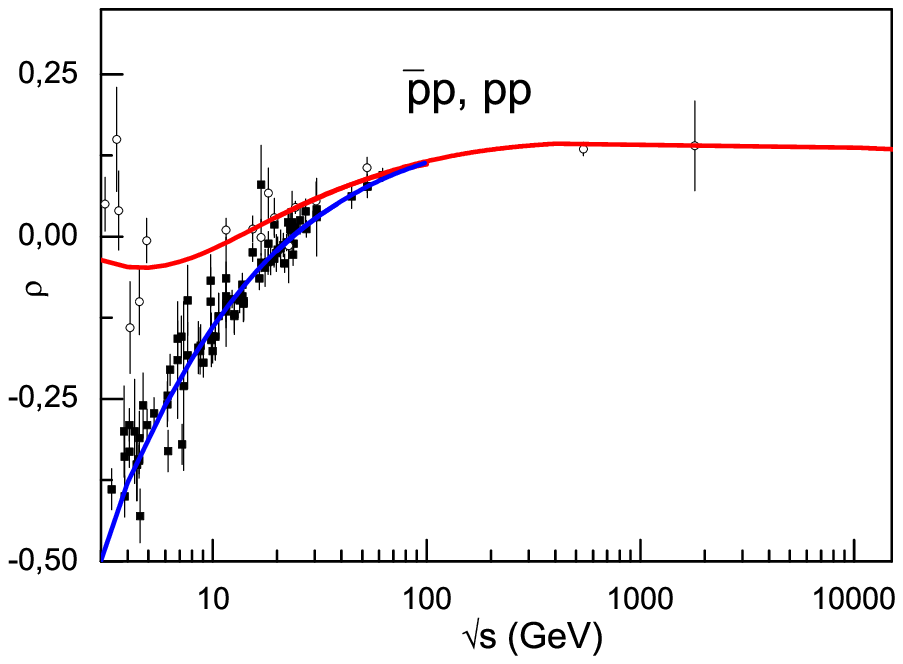}\\
		(a)\hspace{0.45\linewidth}(b)
}
\label{fig:pomeron2}
\caption{
(a) $pp$ and $p\bar p$ total cross sections calculated
from the model, Eq. (2)-(8), and fitted to the data in the range $\sqrt{s}$ = 5 GeV--7 TeV.
(b) Ratio of the real to imaginary part for $pp$ and $p\bar p$ scattering amplitude calculated from the model. The red curve presents the $p\bar p$ calculation, the blue one corresponds to $pp$.}
\end{figure}
\section{Fitting procedure}
\begin{figure}[fig:dif]
\center{
        \includegraphics[width=0.45\linewidth]{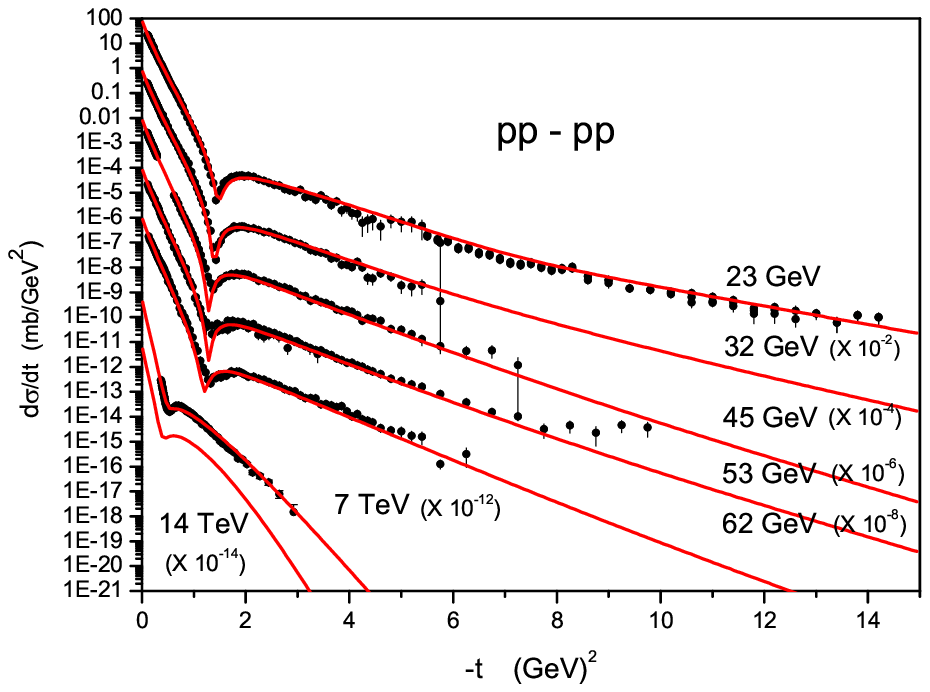}
        \includegraphics[width=0.5\linewidth]{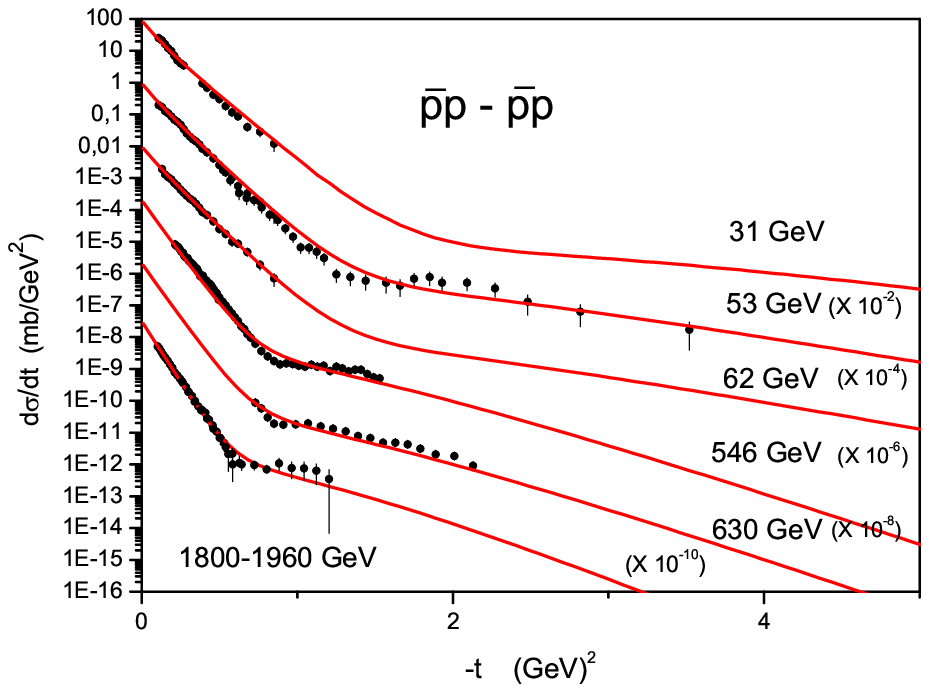}\\
        (a)\hspace{0.45\linewidth}(b)}
\caption{
(a) Differential $pp$ (a) and $pp$ (b) cross sections calculated from the model, Eqs. (2)-(8) with the Odderon term (9)-(11) and fitted to the data in the range $-t$ = 0.1 --- 15~GeV$^2$.
}
\label{fig:dif}
\end{figure}

The model contains (at most) 17 parameters (depending on the choice of the trajectories) to be fitted to about 1200 data points simultaneously in $s$ and $t$.
By a straightforward minimization one has little chances to find the solution, because of possible correlations between different contribution and the parameters, including the $P-f$ and $O-\omega$ mixing and the unbalanced role of different contributions/data points.
To avoid false $\chi^2$ minima, we proceed step-by-step: we first fit the model to the forward data: the
 total cross section and the ratio $\rho=Re A(s,t=0)/Im A(s,t=0)$, starting with the dominant Pomeron contribution  with the sub-leading Reggeons, then we perform the fit for first cone and finally  adding  the Odderon to the whole region of momentum transfer.
 By using the fitted parameters as inputs, we repeat
 the fit with the complete set of the data on elastic $pp$ and $\bar pp$
 elastic scattering differential and total cross sections. The data  compiled in \cite{data} were used in our fitting procedure. The data are: total $pp$ and $\bar{p}p$ cross section measurements spanning energy
 range from 5 to 7 TeV  and to 2.0 TeV, respectively. Another set of
the data are those on the ratio of the real to the imaginary part of the forward amplitude.
These sets contain measurements from both experiments at the Tevatron.
Collection of single-differential elastic cross sections as functions of $t$, measured at different energies were used for the fits.
First of all we check the possible best fit for forward scattering, i.e. fitting the total cross section $\sigma(s)$ and $\rho(s)$ for well the established set of this type of data \cite{data} plus the new measurement at 7 TeV \cite{7pp}.
The quality of the fit is not worse then the standard COMPETE fit \cite{COMPETE}
Although we apply the best global fit (minimal $\chi^2$) as a formal criterion for the valid description, we are primarily interested in the region beyond the first cone, critical for the identification of the assumed Odderon at TeV energies. As mentioned in the Introduction, we perform also a fit to $pp$ data alone, see the previous Section, to see whether the observed dynamics of dip can be reproduced by the Pomeron alone. The contribution to the global $\chi^2$ from tiny effects, such as the small-$|t|$ ``break'' in the first (and second) cone, possible oscillations in the slope of the cone(s) etc. should not corrupt the study of the dynamics in the dip-bump region.
The following kinematical regions and relevant datasets were involved in the fitting procedure:
23, 32, 45, 53, 62 GeV and 7 TeV for $pp$ scattering  \cite{ISR,Amaldi:1979kd,Albrow:1976sv,Breakstone:1984te}
and 31, 53, 62, 546, 630 GeV, 1,8 Tev and 1,96 TeV for $\bar{p}p$ scattering
\cite{Bozzo:1985th,Bernard:1986ye,Amos:1990fw,Abe:1993xx, Royon}. These datasets were compiled in a in~\cite{data}. The differential elastic scattering cross sections were further constrained to cover the momentum transfer range $|t|=$0.05 --- 15 GeV$^2$.
Next, we included in the fit the differential cross sections in first cone chosen, somewhat subjectively for $\left|t\right|\leq 0.5 GeV^{2}$ along with forward data, to determine the remaining parameters of the Reggeons and the Pomeron,$b_{f}, b_{\omega}$ for Reggeons, $a_P, b_P, \alpha_{1P}$ and $\alpha_{2P}$ for the Pomeron,
important in first cone. Among the parameters of the previous fit we fixed the parameters responsible for rise of the total cross section. We performed two series of fits: with linear Pomeron trajectory (${\alpha_{2p}}=0$) and with a nonlinear one (${\alpha_{2p}}\neq 0$). For the grand total of ~ 600 experimental points for the linear trajectory the quality of fit is better for about 70 percent in second case.
It is obvious that the nonlinearity of Pomeron trajectory plays a noticeable role. The presence of a non-negligible curvature in the first cone slope can be clearly seen with the help of the local slope procedure (see, for example, fig.(8) in \cite{JLL}).
    The resulting fits are presented in Table 1. and Fig.~\ref{fig:dif}~.
    Now, for the adequate study of the role of the Pomeron, and especially that of the Odderon outside the first cone,
we must properly choose the parameters with account for their possible correlations.
\begin{table}
\caption {Parameters, quality of the fit and predictions of
 $\sigma_{tot}$ obtained in the whole interval in $s$ and $t$. }
\vskip3mm
\label{tab:fitparam}
\tabcolsep18.2pt
\begin{center}\small
\begin{tabular}{|lcc|} \hline
Parameter & Value & Error\\
\hline
\hline
  $ a_{P}$ & 269 & 5 \\
  \hline
  $b_ {P},GeV^{-2}$ &  6.93 & 0.06 \\
  \hline
  $\alpha_{1P},GeV^{-2}$ &  0.474 & 0.006 \\
  \hline
 $ \alpha_{2P}$ & 0.0060 & 0.0010  \\
  \hline
 $ \delta_p $ & 0.0504 & 0.0031 \\
 \hline
$\epsilon_{P}$ & 0.0167 & 0.0005 \\
\hline
    $ a_{O}$ & 0.160 & 0.006 \\
  \hline
  $ b_{O}, GeV^{-2}$ & 1.79 & 0.09  \\
  \hline
  $\alpha_{1O},GeV^{-2}$ &  0.276 & 0.008 \\
  \hline
 $ \alpha_{2O}$ &  0.339 & 0.017  \\
  \hline
 $ \delta_{O} $ & 0.106 & 0.008 \\
\hline
$\epsilon_{O}$ &-0.223 & 0.032 \\
\hline
  $a_{f}$ & -13.2 & 0.1 \\
  \hline
  $\alpha_{f}$ &  0.790 & 0.004 \\
  \hline
   $ b_f GeV^{-2}$ & 4.24 & 0.15  \\
   \hline
  $ a_{\omega}$ & 8.51 & 0.32 \\
  \hline
  $\alpha_{\omega}$ &  0.473 & 0.012 \\
  \hline
 $ b_{\omega}, GeV^{-2}$ & 15 & fixed  \\
 \hline
$\chi^2/dof$ & 3.55 & \\
\hline
$\sigma_{tot}(7 TeV) $& $98,1\pm 0.1 $ & \\
\hline
$\sigma_{tot} (14 TeV)$ & $111.4\pm 0.1$ & \\
\hline
\end{tabular}
\end{center}
\end{table}
Finally, we note that the best fit to the data does not necessarily
implies the best physical model, but the opposite statement is always true.

\begin{figure}[fig:htbp]
\center{
        \includegraphics[width=0.45\linewidth]{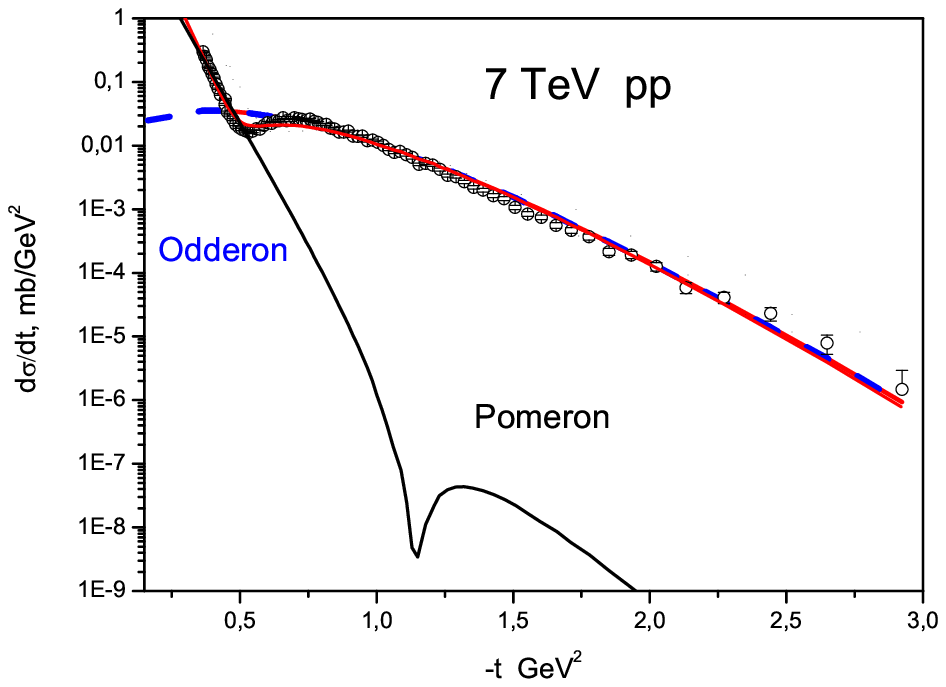}\\
}
\caption{ pp differential cross section at 7 TeV calculated from the model. The solid black line corresponds to the Pomeron contribution, the dashed blue line to that of the Odderon, and solid red line to the differential cross section.}
\label{fig:htbp}
\end{figure}
\section{Conclusions}
A basic problem in studying the Pomeron and Odderon is their identification i.e.
their discrimination from other contributions.
 Although this procedure is model-dependent, we try to do this possibly
 in a general way. We try to answer the important question: where
 (in $s$ and in $t$) and to what extent will be the elastic data from the LHC
dominated by the Pomeron and Odderon contribution? The answer to this question
is of practical importance since, by Regge-factorization, it can
be used in other diffractive processes, such as diffraction
dissociation. It is also of conceptual interest in our definition
and understanding of the phenomenon of high-energy diffraction.
It ensues from our analysis that the dipole model of Pomeron and Odderon unambiguously follows, that  at high (TeV) energies the Pomeron prevails in the first cone, while in the second one the Odderon is dominated, interference of which at least qualitatively describes the dip in $pp$-differential cross section and accordingly  the  plateau in the $p \bar p$
(See Fig~\ref{fig:htbp}~).
The aim of the present paper was
to trace the Pomeron and Odderon contribution under conditions accessible within LHC kinematics. This was feasible due to the simplicity of the model, which has the important property of
reproducing itself (approximately) against unitarity (absorption)
corrections, that are small anyway (for more details see
\cite{ 2,3} and references therein).

We have presented the ``minimal version'' of the DP model. It can be further extended, refined and improved,
while its basic features remain intact.
The anticipated rescaling of the LHC energy down to that of the highest Teavatron energy may provide a definite answer to the questions
concerning the Odderon in $pp$ vs. $\bar pp$ scattering, raised in the present paper.

\section*{Acknowledgments}
We acknowledge fruitful discussions and useful remarks by Tam\'as Cs\"{o}rg\H{o}, L\'aszl\'o Jenkovszky, Denys Lontkovskyi and Mikola Romanyuk. We thank Frigyes Nemes for his help in calculations.

\vfill \eject

\vskip 5 cm
\newpage
\begin{thebibliography}{99}
\bibitem{ISR} A.~Breakstone {\it et al.}, Phys.\ Rev.\ Lett.\  {\bf 54}, 2180 (1985).
\bibitem {7pp} TOTEM Collaboration et al,  EPL 95 (2011) 41001.
\bibitem{VJS} A.N.~Vall, L.L.~Jenkovszky and B.V.~Struminsky, EChAYa (Russian translation: PEPAN) {\bf 19} (1988) 180.
\bibitem{C-Y} T.T.~Chou, and C.N.~Yang, Phys .Rev.Lett. {\bf 20} (1968) 1615.
\bibitem {JLL} L.L.Jenkovszky, A.I.Lengyel, D.I.Lontkovskyi, Int. J. Mod. Phys. A 26 (2011) 4577.
\bibitem{Godizov}A.A.Godizov. Models of elastic diffractive scattering to falsity at the LHC. ArXiv1203.6013v1 [hep-ph] 27 mar 2012.
\bibitem{Nicolescu2}Basarab Nicolescu. Recent advances in Odderon physics. ArXiv9911334v1 [hep-ph] 12 nov 1999.

\bibitem{Nicolescu}Basarab Nicolescu. The Odderon at  RHIC and LHC.  ArXiv0707.2923v1 [hep-ph] 19 jul 2007.

\bibitem{reviews} L.~Jenkovszky, Fortschritte der Physik, {\bf 34} (1986); L.~Jenkovszky, Rivista Nuovo Cim. {\bf 10} (1987) 1;
    L.~Jenkvoszky, EChAYa (Egl. translation: PEPAN) {\bf 34} (2003) 1196.

\bibitem {7tev} TOTEM Collaboration G.Antchev et al. Europhys.Lett {\bf 96} 21002 (2011).

\bibitem{DL} S.~Donnachie and P.~Landshoff, Phys. Lett. {\bf B 123} (1983) 345; Nucl. Phys. {\bf 267} (1985) 690.
\bibitem{review} R.~Fiore {\it et al.}
Int. J. Mod. Phys., A24 (2009) 2551-2559, arXiv:hep-ph/0812.0539.



\bibitem{2} P.~Desgrolard, M.~Giffon, L.L.~Jenkovszky, Z. Phys. C
{\bf 55}(1992) 643.
\bibitem{3} R.J.M.~Covolan, P.~Desgrolard, M.~Giffon, L.L.~Jenkovszky and E.~Predazzi,
Z.Phys. C {\bf 58} (1993) 109.
\bibitem{KLT} K.~Kontros, A.~Lengyel, and Z.~Tarics, {\it $pp$ and $p\bar p$ elasctic scattering in a multipole Pomeron model}, hep-ph/0011398.



\bibitem{Cohen} G.~Cohen-Tannoudji {\it et al.} Lettere Nuovo Cim. {\bf 5} (1972) 957.

\bibitem{Amaldi:1979kd} U.~Amaldi and K.R.~Schubert,  Nucl.\ Phys.\  B {\bf 166}, 301 (1980);
\bibitem{Albrow:1976sv} M.G.~Albrow {\it et al.},  Nucl.\ Phys.\  B {\bf 108}, 1 (1976);
\bibitem{Breakstone:1984te} A.~Breakstone {\it et al.}, Nucl.\ Phys.\  B {\bf 248}, 253 (1984);


\bibitem{data} http://qcd.theo.phys.ulg.ac.be/$\sim$cudell/;\\
http://qcd.theo.phys.ulg.ac.be/$\sim$cudell/DATA.html/;\\
http://www.theo.phys.ulg.ac.be/$\sim$cudell/data/;\\
http://pdg.lbl.gov/2002/;

\bibitem{COMPETE} J.R. Cudell et al., Phys. Rev. D65, 074024 (2002) [arXiv:hep-ph/0107219]; J.R. Cudell et al., Phys. Rev. Lett. 89 (2002) 201801; [arXiv:hep-ph/0206172].

\bibitem{Bozzo:1985th} M.~Bozzo {\it et al.}, Phys.\ Lett.\  B {\bf 155}, 197 (1985);
\bibitem{Bernard:1986ye} D.1.~Bernard {\it et al.},  Phys.\ Lett.\  B {\bf 171}, 142 (1986);
\bibitem{Amos:1990fw} N.A.~Amos {\it et al.}, Phys.\ Lett.\  B {\bf 247}, 127 (1990);
\bibitem{Abe:1993xx} F.~Abe {\it et al.}, Phys.\ Rev.\  D {\bf 50}, 5518 (1994);
\bibitem{Royon} C.Royon, D0 results on diffraction, 14thWorkshop on Elastic and Diffractive Scattering December 15-21 2011, Qio Nhon  Vietnam.


\end{thebibliography}
\end{document}